\documentclass[twoside]{dis09}
\usepackage[latin1]{inputenc}
\usepackage[dvips]{graphicx,epsfig,color}
\usepackage{wrapfig,rotating}
\usepackage{amssymb,amsmath,array}

\usepackage[english]{babel}
\usepackage{epsfig,graphicx}
\usepackage{amsbsy}
\usepackage{amsfonts}
\usepackage{pgf}
\usepackage{xcolor}
\usepackage{verbatim}

\newcommand{\tw}{\textwidth}
\newcommand{\be}{\begin{equation}\nonumber}
\newcommand{\ee}{\end{equation}}
\newcommand{\bea}{\begin{eqnarray}}
\newcommand{\eea}{\end{eqnarray}}

\newcommand{\bc}{\begin{center}}
\newcommand{\ec}{\end{center}}

\newcommand{\bi}{\begin{itemize}}
\newcommand{\ei}{\end{itemize}}

\definecolor{red}{rgb}{1,0,0}
\definecolor{green}{rgb}{0.3,0.6,0.3}
\definecolor{blue}{rgb}{0,0,1}
\definecolor{darkgreen}{rgb}{0,0.39,0.00}

\begin{document}
\title{Towards small-$x$ resummed DIS phenomenology}

\author{Juan~Rojo$^1$, 
Guido~Altarelli$^2$,
Richard~D.~Ball$^3$ and Stefano~Forte$^1$, \\
$^1$Dipartimento di  Fisica, Universit\`a di
Milano and  INFN, Sezione di Milano,\\ Via Celoria 16, I-20133 Milan, Italy\\
$^2$Dipartimento di Fisica ``E.Amaldi'', 
Universit\`a Roma Tre and INFN, Sezione di Roma Tre\\
Via della Vasca Navale 84, I--00146 Roma, Italy, \\ 
 CERN, Department of Physics, Theory Division, CH-1211 Gen\`eve 23,
 Switzerland \\
$^3$School of Physics, University of
Edinburgh,  Edinburgh EH9 3JZ, Scotland
} 


\maketitle

\begin{abstract}
We report on recent progress towards quantitative phenomenology
of small $x$ resummation of deep--inelastic  
structure functions. We compute small $x$ resummed $K$--factors
with realistic PDFs and estimate their impact
in the HERA kinematical region.
These $K$--factors, which match smoothly to
the fixed order NLO results, approximately reproduce the effect of a
small $x$ resummed PDF analysis.
 Typical
 corrections   are found to be of the
same order as the  NNLO ones, that is,
a few percent, but with opposite
sign. These results imply that resummation corrections could be relevant
for a global PDF analysis, especially 
with the very precise combined  HERA dataset.
\end{abstract}

\paragraph{Small $x$ resummation in the LHC era}

The so-called small $x$ regime of QCD is the kinematical region in
which hard scattering processes happen at a center-of-mass
energy which is much larger than the characteristic hard scale
of the process. An understanding of strong interactions in this region
is therefore necessary to do physics at high--energy colliders. 
In this sense, HERA was
the first small $x$ machine, and LHC is going to be even more of a
small $x$ accelerator. 

As is by now well known, perturbative corrections become large at
small $x$. Due to the accidental vanishing of some
coefficients, the leading large corrections cannot be seen in LO and
NLO splitting functions; however, the first subleading correction can
already be seen in the NNLO splitting functions which have been
computed recently, as well as in NNLO coefficient functions: they are
large enough to make recent NNLO parton fits unstable at small
$x$~\cite{Martin:2009iq}. 

This suggests dramatic effects from yet higher
orders, so the success of NLO perturbation theory at HERA, as
demonstrated by the scaling laws it predicts, has been
for a long time very hard to explain.
In the last several years this situation has been clarified~\cite{Altarelli:2003hk,Altarelli:2005ni,Altarelli:2008aj,Ciafaloni:2003kd,Ciafaloni:2007gf},
showing that, once all relevant large terms are included, the effect of
the resummation of terms which are enhanced at small $x$ is
perceptible but moderate --- comparable in size to typical NNLO fixed
order GLAP corrections in the HERA region. 
A recent status report on small $x$ resummation, including a comparison
of various approaches and a more complete list of references, can
be found in
 Ref.~\cite{Dittmar:2009ii}.

Recently, in Ref.~\cite{Altarelli:2008aj} 
 a full small $x$ resummation
including quarks and the resummation of deep-inelastic
coefficient functions was presented, so that resummed 
expressions for deep-inelastic structure
functions can be obtained. 
Furthermore, the resummation of hard partonic cross sections
has been performed in several LHC processes such as 
heavy quark production~\cite{Ball:2001pq}, 
Higgs production~\cite{Marzani:2008az,Marzani:2008ih},
Drell-Yan~\cite{Marzani:2008uh,Marzani:2009hu} 
and prompt photon production~\cite{Diana:2009xv}.
This will enable fully resummed phenomenology. In
this contribution we present a first step in this direction.

\paragraph{Small $x$ resummed $K$--factors}

\begin{figure}[htb]
\begin{center}
\includegraphics[width=0.49\tw]{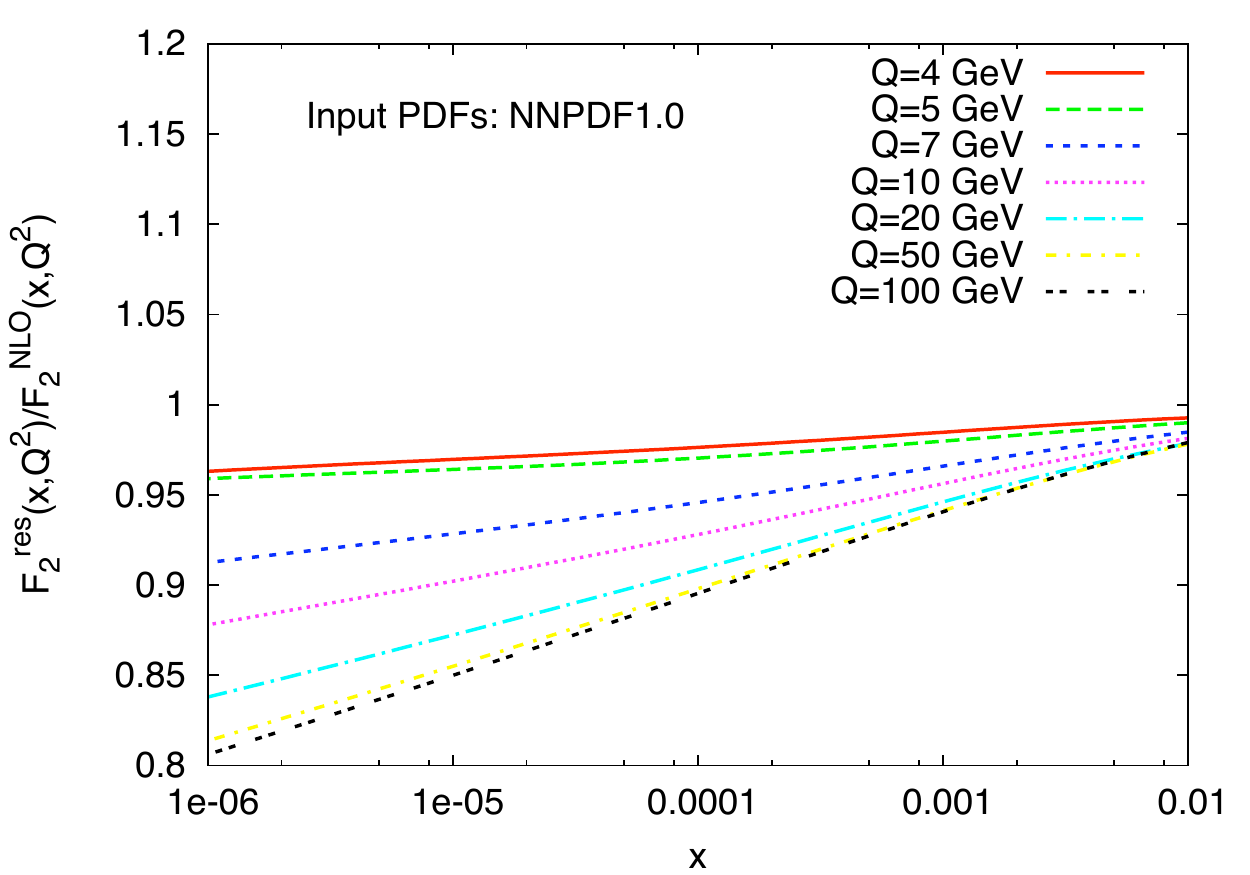}
\includegraphics[width=0.49\tw]{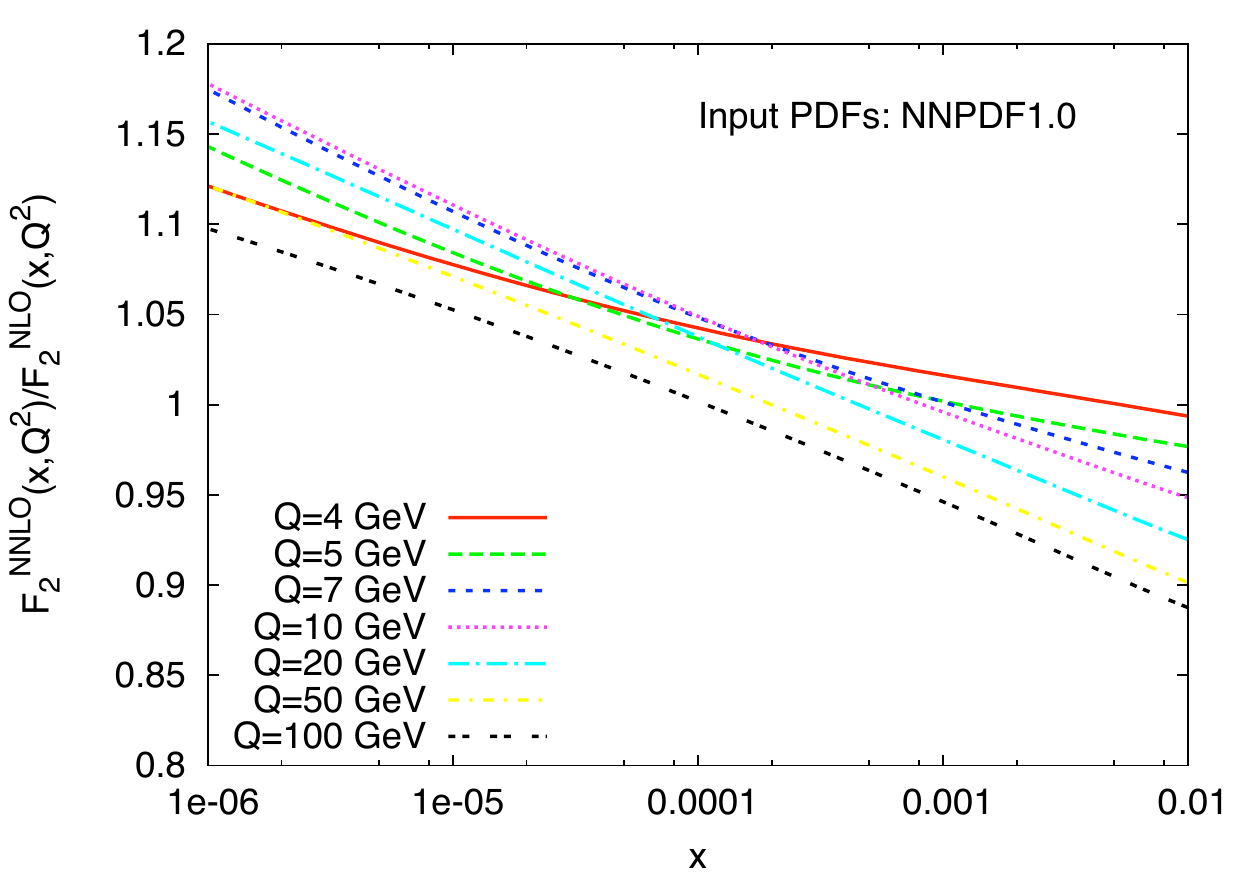}
\includegraphics[width=0.49\tw]{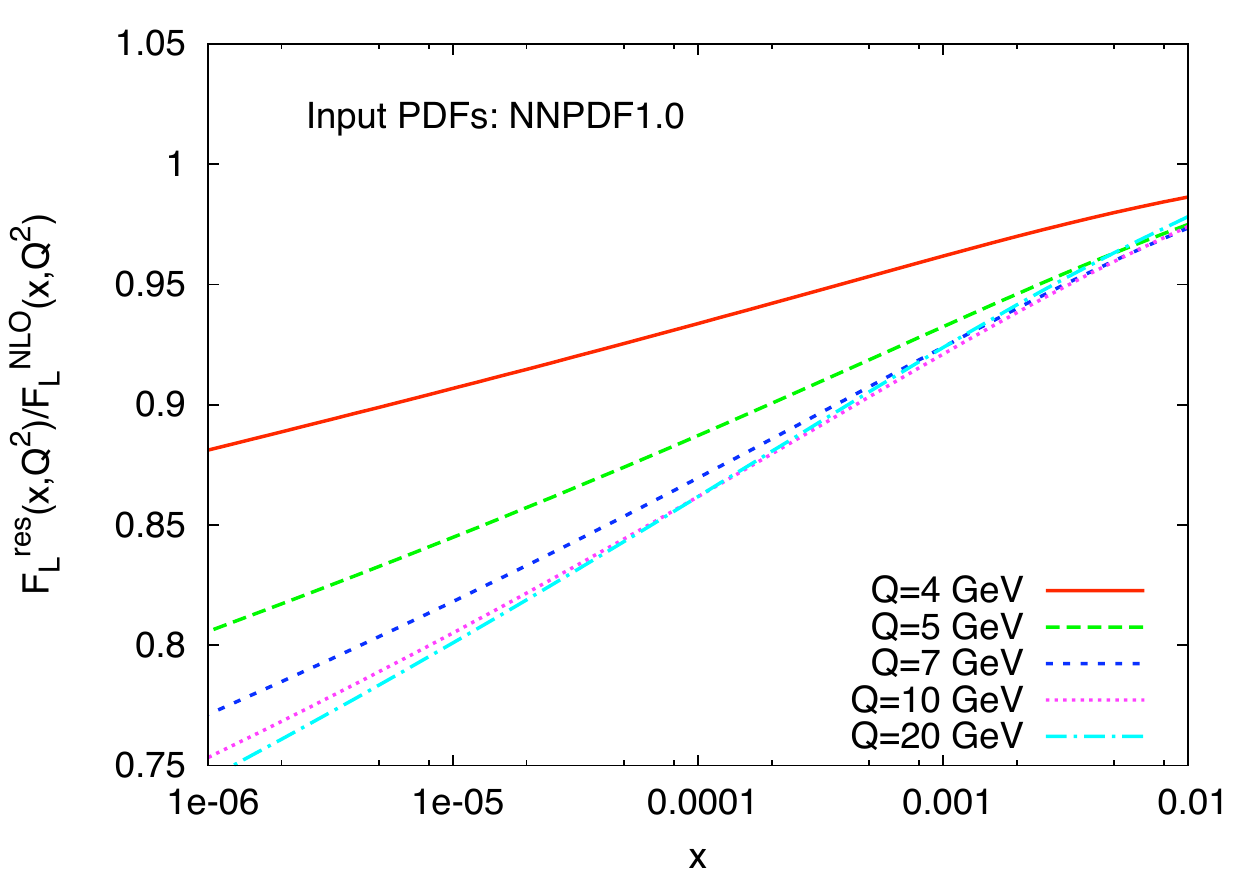}
\includegraphics[width=0.49\tw]{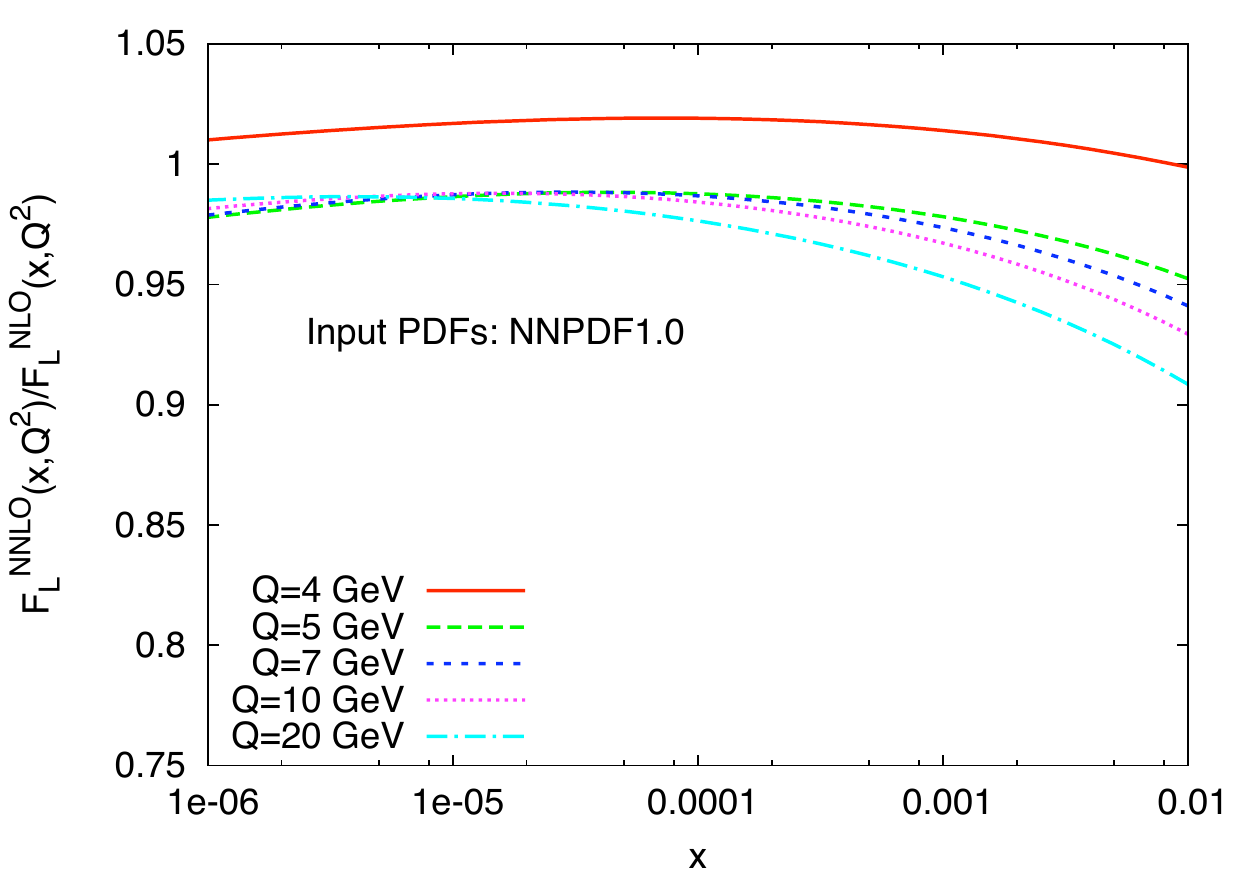}
\caption{\small Upper left: the small $x$ resummed $K$--factors for
$F_2(x,Q^2)$
for various values of $Q$.
Upper right: analogous $K$--factors for the NNLO case.
Lower left: the small $x$ resummed $K$--factors for
$F_L(x,Q^2)$
for various values of $Q$.
lower right: analogous $K$--factors for the NNLO case.
In all cases,  the matching scale has been
taken to be $Q_0=3$ GeV and the input PDF set is NNPDF1.0.
\label{fig_kfact}}
\end{center}
\end{figure}

Using the results
of Ref.~\cite{Altarelli:2008aj}, it is now
possible to determine resummed predictions for deep--inelastic
structure functions. Eventually, these should be combined with resummed
expressions for various parton-level cross-sections in order
to determine parton distributions at the resummed level, and use
them for a fully resummed treatment of hard processes.

However, as a first step in this program, it is convenient
to provide a qualitative estimate of the impact of small $x$
resummation on DIS phenomenology.
Such an estimate can
be obtained by first computing the structure functions $F_2$ and $F_L$
with a given set of NLO PDFs, and then 
assuming that the structure functions 
are kept fixed at some scale $Q_{0}$: 
this is then enough to determine the resummed
singlet quark and gluon distributions at that scale. 
These quark and gluon distributions are close to 
those which would
be obtained if PDFs were determined from a fit to DIS data mostly
clustered around $Q_0$. They can thus be used to compute 
observables at
any other scale.

We have used them to
 recompute the structure functions $F_2$ and $F_L$ in the resummed 
formalism and we have then determined $K$--factors as ratios
 of  the results from the whole procedure at the resummed and at the 
NLO levels, respectively. In the
following, we will show results obtained with the choice $Q_0=3$ GeV,
which roughly corresponds to 
the typical scale of the smallest $x$ HERA data used in a parton fit.
Different choices of $Q_0$ lead to rather different results,
 so that this $K$--factor  approach can only be used to get a first
 qualitative feeling for the phenomenological impact of resummation.

$K$--factors thus obtained for the $F_2(x,Q^2)$
and $F_{L}(x,Q^2)$ structure functions
are shown in Fig.~\ref{fig_kfact}.
In order to allow a more direct comparison, 
NNLO $K$--factors have been computed 
in the same way as the NLO small $x$ resummed ones.
As one can observe, the impact of the resummation at the ``HERA scale''
$Q_0$
is comparable to that of
NNLO corrections, but it goes in the opposite direction: it tends to
suppress the starting PDFs while the NNLO tends to
enhance them. Note that in the small $x$ resummed case asymptotic
freedom is at work: for large values of $Q$, the $K$--factors become
independent of its precise value.

The determination of $K$--factors requires the use of an input
parton set. The $K$--factors shown in
 Fig.~\ref{fig_kfact} have been computed using the
 NNPDF1.0~\cite{Ball:2008by} parton set\footnote{
The NNPDF approach to parton distributions~\cite{DelDebbio:2007ee,
Ball:2008by,Rojo:2008ke,nnpdf12} provides 
sets of PDFs with faithful uncertainty estimation thanks
to a combination of artificial neural networks as
unbiased interpolations and Monte Carlo methods for
robust error estimation and propagation.}. We have explicitly checked that
different choices, for example using as input PDFs not the central
NNPDF1.0 set but a set of PDFs which roughly sit on the associated
$\pm 1$-$\sigma$
PDF uncertainty band, have a negligible impact on the procedure.

As already mentioned, from Fig.~\ref{fig_kfact} the dominant
qualitative feature of the $K$--factors is that resummation leads to
a suppression of the structure functions $F_2$ and $F_L$ at small
values of
$x$. Note also the smooth matching with the GLAP NLO result, since
the $K$--factor tends to 1 as expected at moderate and large-$x$. 

It is important to understand the 
meaning of these $K$--factors, which increasingly deviate 
from one at larger $Q^2$, and not at low $Q^2$ as one
might expect. Indeed, these $K$--factors provide the change  in
prediction, from NLO to the resummed level, for structure functions
 at large $Q^2$, assuming that the 
structure functions at the low scale $Q_0^2$ are given and fixed. 
As such, they provide a model for 
the change due to resummation in prediction at LHC scales, 
when current HERA data are used to determine parton 
distributions at small $x$.

\paragraph{Small $x$ resummed phenomenology}

\begin{figure}[htb]
\begin{center}
\includegraphics[width=0.77\tw]{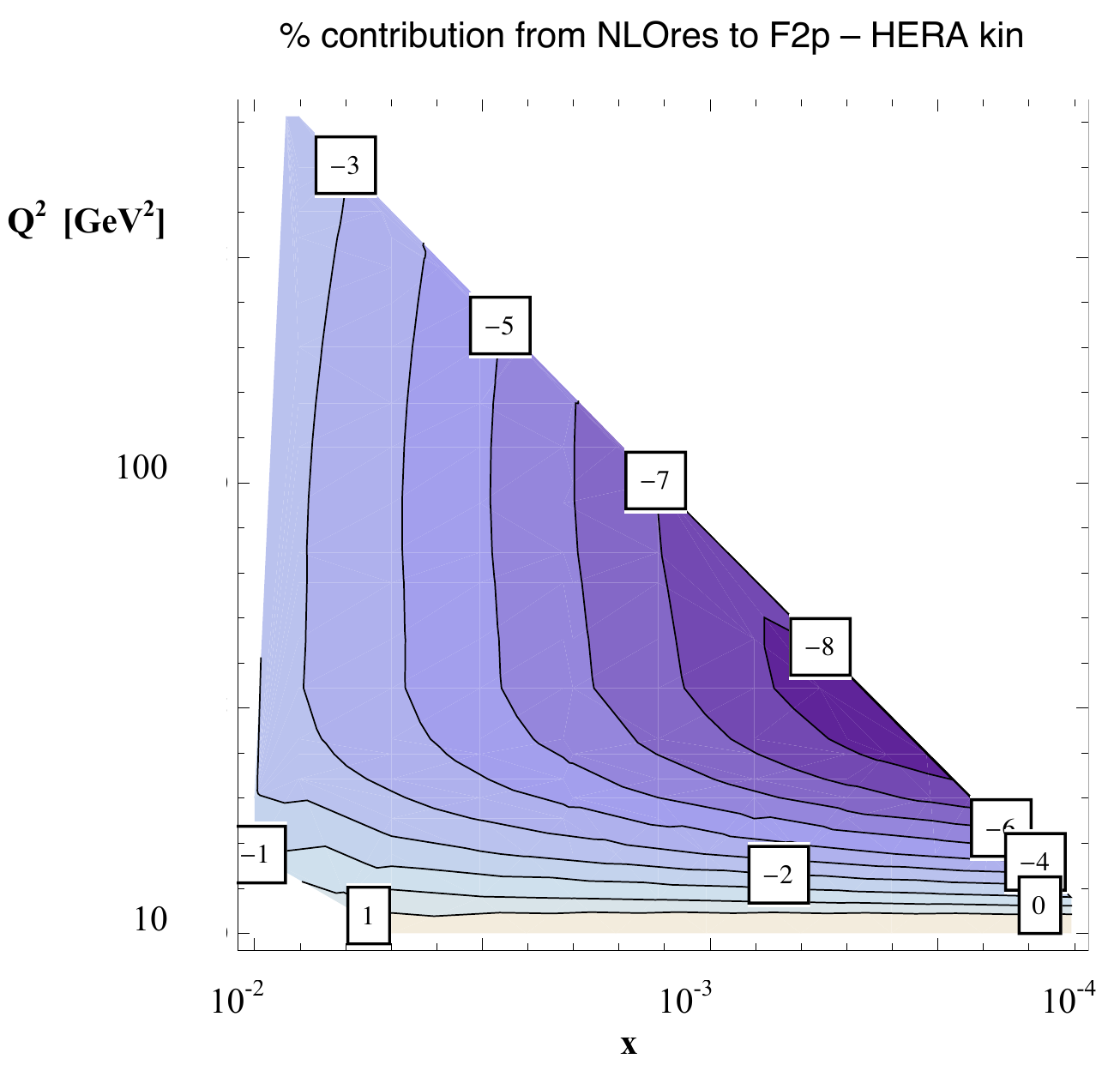}\\
\end{center}
\caption{\small The impact of small $x$ resummation on the
kinematics of published $F_2(x,Q^2)$ HERA data. The various
contours show the relative percent difference between the
NLO and the small $x$ resummed computations.\label{fig:resummedHERA}}
\end{figure}

Once the small $x$ resummed $K$--factors have been
computed, they can be used to obtain resummed 
predictions for structure functions using any input NLO
set of PDFs. 
In order to get an estimate of the quantitative impact of
small $x$ resummation in the HERA region, in Fig.~\ref{fig:resummedHERA}
we show the relative differences between the NLO computation and the
small $x$ resummed one in the $(x,Q^2)$ kinematics of the published
HERA data. As before, the matching scale has been taken to
be $Q_0=3$ GeV and the input PDF set is NNPDF1.0. We observe that the typical
effect is a negative difference (the resummation decreases $F_2$)
of the order of a few percent. Since the upcoming combined HERA
data set will have an accuracy of $\sim 1\%$, this implies that the
effects of small $x$ resummation should be relevant in a global PDF
analysis.

\begin{figure}[htb]
\begin{center}
\includegraphics[width=0.80\tw]{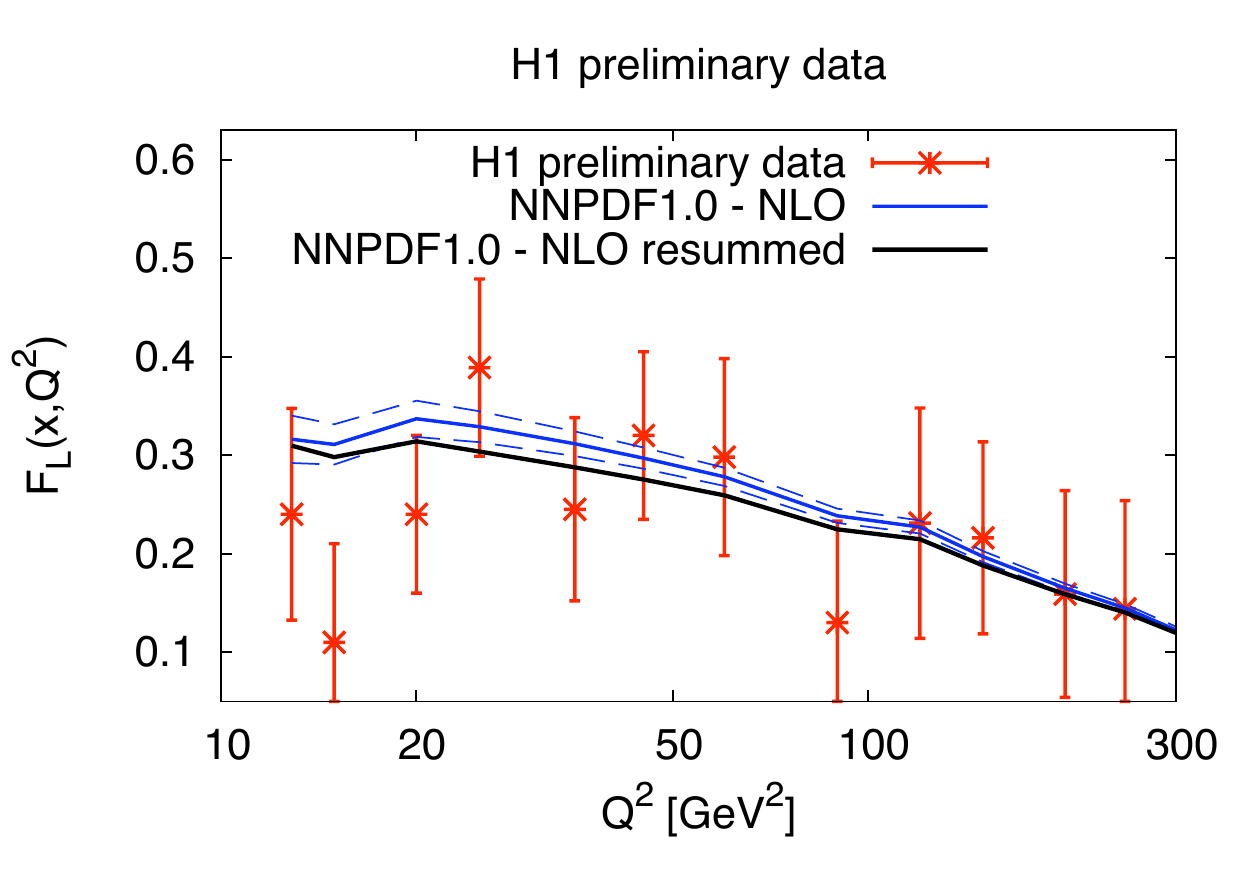}\\
\end{center}
\caption{\small Comparison between the H1 measurent of
$F_{L}(x,Q^2)$~\cite{h1fl} 
and the NLO and small $x$ resummed predictions computed with
the NNPDF1.0 parton set.
\label{fig:fl}}
\end{figure}

\begin{figure}[htb]
\begin{center}
\includegraphics[width=0.80\tw]{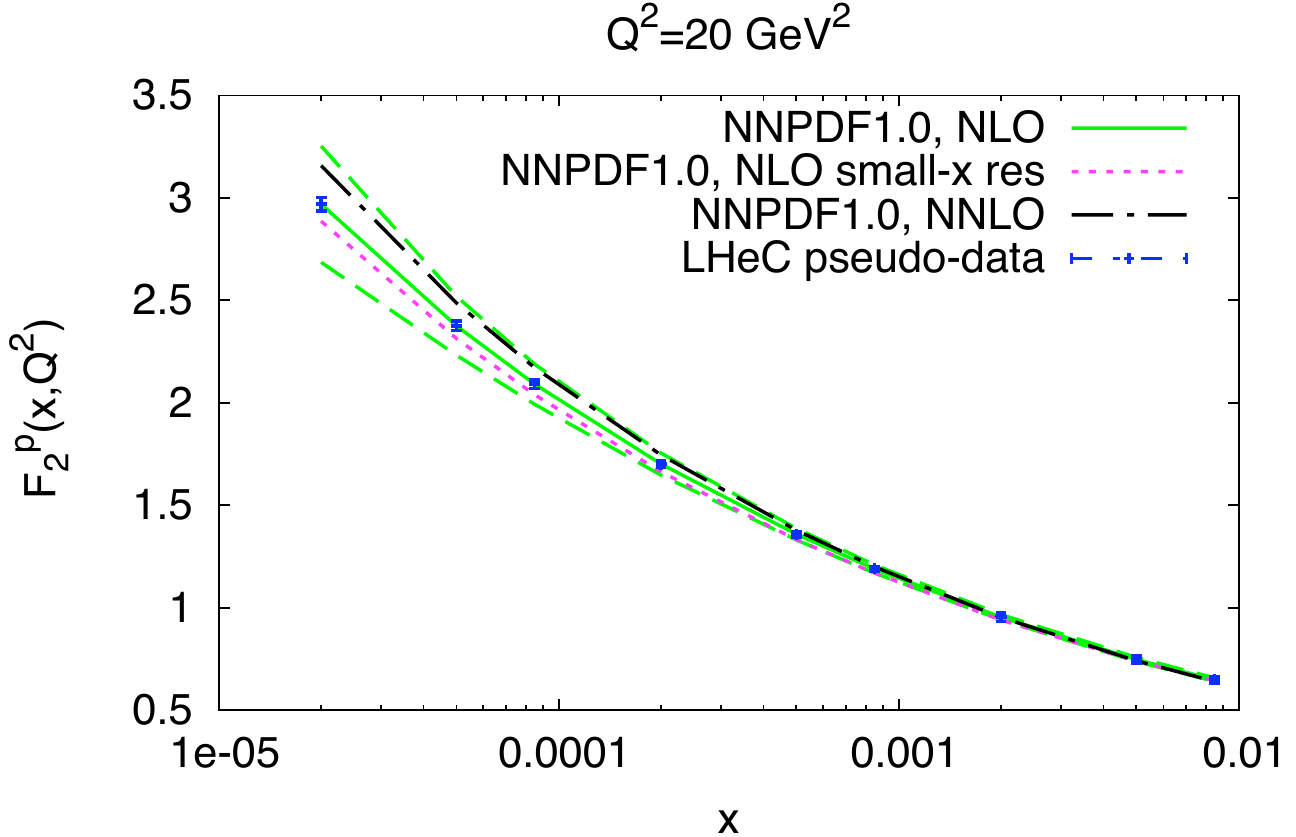}\\
\includegraphics[width=0.05\tw]{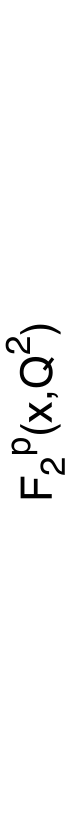}
\includegraphics[width=0.75\tw]{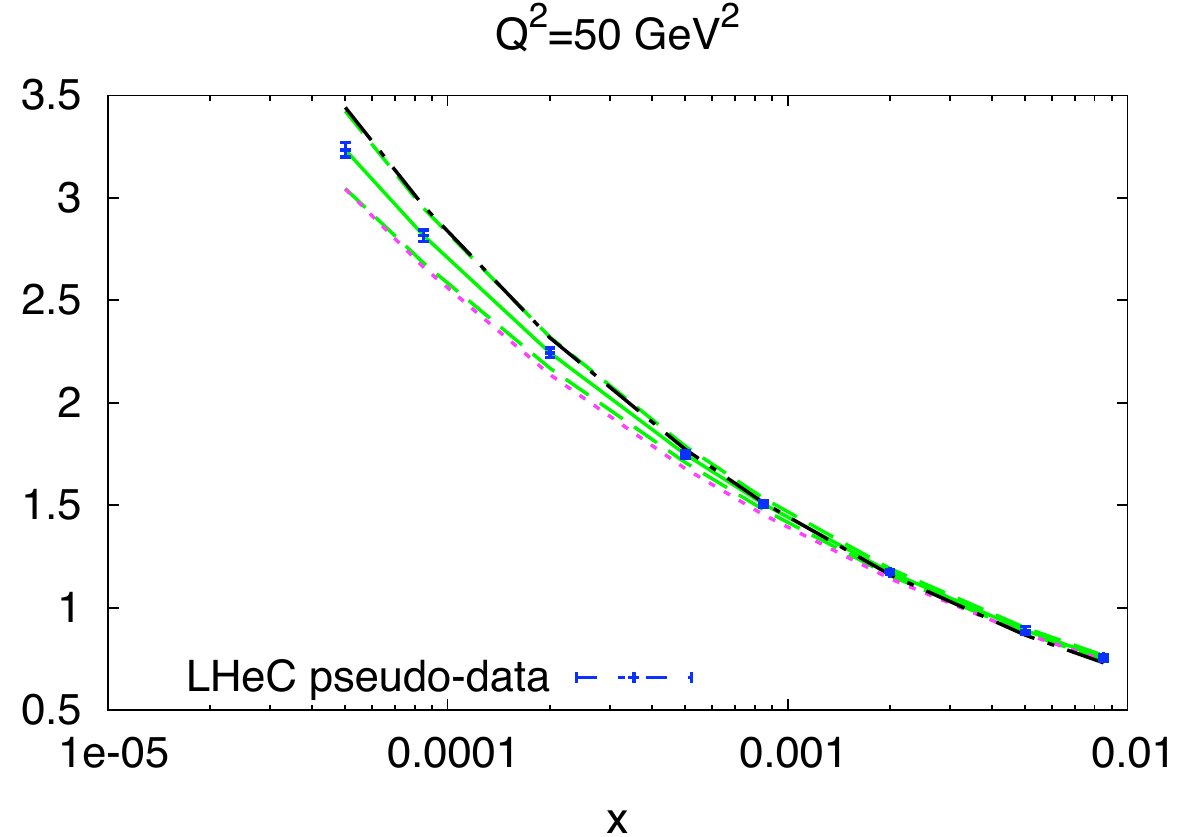}
\end{center}
\caption{\small A comparison of various
approximations to linear low-$x$ QCD for $F_2$
at the LHeC: the NNPDF1.0
prediction which includes PDF
uncertainties and the
NNPDF1.0 result corrected with the
NNLO and small $x$ resummed 
$K$--factors. The expected 
 experimental precision at the LHeC
is also shown for illustration. The upper
plot corresponds to $Q^2=$ 20 GeV$^2$, while
the lower plot to  $Q^2=$ 50 GeV$^2$.  \label{f2resum}}
\end{figure}

%
%


As another example of the applications of the resummed $K$--factors,
we show in Fig.~\ref{fig:fl} a comparison between H1 data on
the longitudinal structure function $F_L(x,Q^2)$~\cite{h1fl} 
and the NNPDF1.0
prediction both at NLO and with the small $x$ resummed $K$--factors.
Interestingly, the
suppression due to the resummation is generally
larger than the one--sigma band due to PDF uncertainties, although the
statistical accuracy of the measurements is still not enough to
provide any discrimination power.

The impact of resummation on partonic cross sections
for LHC signal, background or standard candle
processes such as respectively Higgs~\cite{Marzani:2008az}, 
prompt photon~\cite{Diana:2009xv}
 and Drell-Yan~\cite{Marzani:2008uh} are now 
known to be qualitatively similar
to the impact on deep-inelastic coefficient 
functions~\cite{Ball:2007ra,Marzani:2009hu}. Therefore, it is clear
from the results presented above that resummation is necessary for LHC 
phenomenology
at the percent level of accuracy, typical of NNLO computations.

On a more speculative level, resummation would also be very relevant for
deep-inelastic scattering at a high energy electron-hadron collider
based on the LHC, the so-called LHeC~\cite{Dainton:2006wd}.
Indeed,
in Ref.~\cite{Rojo:2009ut} it was show
how the 
structure function $F_2$  in the LHeC kinematics varies on
applying either resummed or NNLO $K$--factors to the NLO prediction
from the NNPDF1.0 parton set. As can be seen
in Fig.~\ref{f2resum}, the expected accuracy
of such a machine would clearly discriminate between the small $x$
resummed and NNLO cases, since at such high energies the difference
between the two predictions is larger than both the expected
PDF uncertainty and the accuracy of experimental data.

\paragraph{Outlook}
This contribution summarizes ongoing work towards
small $x$ resummed phenomenology of deep-inelastic scattering.
In summary, the impact of
small $x$ resummation is typically 
as large as that of NNLO corrections in the HERA region, 
and even larger
at higher collider energies, 
with the resummed results generally lying
outside the PDF uncertainty bands.

Our results indicate the small $x$ resummed corrections
could be disentangled with the ultimate precision of
HERA data. However,  a fully small $x$ resummed global PDF analysis is
unavoidable in order to quantify the modifications in the
PDFs due to small $x$ resummation, and to propagate these
modifications into relevant observables at the LHC.

Such a programme will be necessary in order to achieve phenomenology at the 
percent level for many LHC signal, background and standard candle 
processes. It would be even more important for phenomenology
at a future LHeC electron-proton collider, and mandatory for the treatment
of extremely high-energy scattering processes, such as those induced by
Ultra-High Energy cosmic neutrinos, which are currently under investigation.

\paragraph{Acknowledgments}
J.~R. acknowledges the hospitality
of the CERN TH Division where part of this work was completed.

\begin{footnotesize}



\providecommand{\href}[2]{#2}\begingroup\raggedright\endgroup

\end{footnotesize}


\end{document}